\documentclass[11pt,leqno]{article}

\usepackage{vmargin}
\setmarginsrb{1in}{1in}{1in}{1in}{0mm}{0mm}{0mm}{7mm}

\usepackage{amsmath,amsthm}
\usepackage{amssymb}
\usepackage{xspace}
\usepackage{color,enumerate}
\usepackage{tikz}
\tikzstyle{every picture} = [>=latex]

\usepackage{todonotes}

\usepackage[czech,english]{babel}

\newtheorem{theorem}{Theorem}[section]
\newtheorem{lemma}[theorem]{Lemma}

\newtheorem{corollary}[theorem]{Corollary}
\newtheorem{proposition}[theorem]{Proposition}

\theoremstyle{definition}
\newtheorem{definition}[theorem]{Definition}

\newtheorem{remark}[theorem]{Remark}

\newcommand{\naturals}{\mathbb{N}}

\newcommand{\AAA}{\mathcal{A}}
\newcommand{\PPP}{\mathcal{P}}

\newcommand{\bigO}{\mathcal{O}}

\newcommand{\card}[1]{{|#1|}}		

\newcommand{\SB}{\{\,}%
\newcommand{\SM}{\;{|}\;}%
\newcommand{\SE}{\,\}}%
\newcommand{\Card}[1]{\card{#1}}

\newcommand{\incom}{\parallel}
\newcommand{\width}{\textup{width}}
\newcommand{\elmax}{\mbox{\it`max'}\xspace}
\newcommand{\elmin}{\mbox{\it`min'}\xspace}
\newcommand{\elup}{\mbox{\it`up'}\xspace}
\newcommand{\eldown}{\mbox{\it`down'}\xspace}

\newcommand{\tuple}[1]{\langle{#1}\rangle}  

\renewcommand{\phi}{\varphi}
\newcommand{\cG}{\mathcal{G}}

\newcommand{\defparproblem}[4]{
  \vspace{4pt}
\noindent\fbox{
  \begin{minipage}{0.96\columnwidth}
  \begin{tabular*}{\textwidth}{@{\extracolsep{\fill}}lr} #1  & {\bf{Parameter:}} #3 \\ \end{tabular*}
  {\bf{Input:}} #2  \\
  {\bf{Question:}} #4
  \end{minipage}
  }
  \vspace{4pt}
}

\usepackage{amstext}

\title{FO Model Checking on Posets of Bounded Width}
\def\inst#1{${}^{#1}\,\vbox to 1.9ex{\vfill}$}
\def\email#1{{\small\tt#1}}
\author{Jakub Gajarsk\'{y}\inst{3} \and Petr Hlin\v{e}n\'{y}\inst{3} \and
  Daniel Lokshtanov\inst{1} \and Jan Obdr\v{z}\'{a}lek\inst{3} \and 
  Sebastian Ordyniak\inst{3} \and
  M. S. Ramanujan\inst{1} \and Saket Saurabh\inst{1,2}
\\[2ex]\inst1\small
    University of Bergen,
    Bergen, Norway, \\
    \email{\{daniello,Ramanujan.Sridharan\}@ii.uib.no}
    \and\inst2\small
    The Institute of Mathematical Sciences, 
    Chennai, India, \\
    \email{saket@imsc.res.in}
    \and\inst3\small
    Faculty of Informatics, Masaryk University, 
    Brno, Czech Republic. \\
    \email{\{gajarsky,hlineny,obdrzalek,ordyniak\}@fi.muni.cz}
\\~
}
\date{}

\begin{document}
\maketitle

\begin{abstract}
Over the past two decades the main focus of research into first-order (FO)
model checking algorithms have been sparse relational 
structures---culminating in the FPT-algorithm by Grohe, Kreutzer and Siebertz for
FO model checking of nowhere dense classes of graphs [STOC'14], with dense
structures starting to attract attention only recently.  Bova, Ganian and
Szeider [LICS'14] initiated the study of the complexity of FO model checking
on partially ordered sets (posets).  Bova, Ganian and Szeider showed that
model checking {\em existential} FO logic is fixed-parameter tractable (FPT)
on posets of bounded width, where the width of a poset is the size of the
largest antichain in the poset.  The existence of an FPT algorithm for
general FO model checking on posets of bounded width, however, remained open.  We resolve this
question in the positive by giving an algorithm that takes as its input an
$n$-element poset~${\cal P}$ of width $w$ and an FO logic formula $\phi$, and
determines whether $\phi$ holds on ${\cal P}$ in time $f(\phi,w)\cdot n^2$.
\end{abstract}

\section{Introduction}
Algorithmic meta-theorems are general algorithmic results applying to a whole range of problems, rather than just to a single problem alone. Such results are some of the most sought-after in algorithmic research. 
Many prominent algorithmic meta-theorems are about {\em model checking};
such theorems state that for certain kinds of logic ${\cal L}$, and all
classes ${\cal C}$ that have a certain structure, there is an algorithm that
takes as an input a formula $\phi \in {\cal L}$ and a structure $S \in {\cal
C}$ and efficiently determines whether $S \models \phi$. Here $S \models
\phi$ is read as ``$S$ models $\phi$'' or ``$\phi$ holds on $S$''. Examples
of theorems of this kind include the classic theorem of
Courcelle~\cite{Courcelle90}, as well as a large body of work on model checking first-order (FO) logic~\cite{bgs14,DawarGK07,DvorakKT10,FlumG01,FrickG01,GHKOST13,ghoo14,gks14,Seese96}. 

Most of the research on algorithms for FO model checking has focused on
graphs. On general graphs there is a naive brute-force algorithm that takes
as an input an $n$-vertex graph $G$ and a formula $\phi$ and determines whether $G \models \phi$ in time $n^{O(|\phi|)}$ by enumerating all the possible ways to instantiate the variables of $\phi$. 
On the other hand, the problem is PSPACE-complete (see
e.g.~\cite{Gradeletal05}) and encodes the {\sc Clique} problem, thus it
admits no algorithm with running time $f(\phi)n^{o(|\phi|)}$ for any
function $f$~\cite{LokshtanovMS11}, assuming the Exponential Time
Hypothesis~\cite{ImpagliazzoPZ01} (ETH). Thus, assuming the ETH the naive
algorithm is the best possible, up to constants in the
exponent. Furthermore, FO model checking remains PSPACE-complete on any
fixed graph containing at least two vertices (again, see~\cite{Gradeletal05}). 
Hence, it is futile to look for restricted classes of graphs in which FO model
checking can be done in polynomial time without restricting $\phi$.
Therefore, research has focused
on obtaining algorithms with running time $f(\phi)n^{O(1)}$ on restricted
classes of graphs and other structures.  Algorithms with such a running time are
said to be {\em fixed parameter tractable} (FPT) parameterized by $\phi$. 
Even though FPT algorithms are not polynomial time algorithms, due to
unlimited $f$, they significantly outperform brute-force.

The parameterized complexity of FO model checking on {\em sparse} graph
classes is now well understood. 
In 1994 Seese~\cite{Seese96} showed an FPT algorithm for FO model checking
on graphs of bounded degree.  Seese's algorithm was followed by a long line
of work~\cite{FrickG01,FlumG01,DawarGK07,DvorakKT10} giving FPT algorithms
for progressively larger classes of sparse graphs, culminating in the FPT
algorithm of Grohe, Kreutzer and Siebertz~\cite{gks14} on any nowhere dense
graph classes.
To complement this, Kreutzer~\cite{Kreutzer11} and Dvo\v{r}\'{a}k et
al.~\cite{DvorakKT10} proved that if a class ${\cal C}$ closed under taking
subgraphs is not nowhere dense, then deciding first-order properties of
graphs in ${\cal C}$ is not fixed-parameter tractable unless FPT=W[1]
(a complexity-theoretic collapse which is considered to be unlikely).  
Hence, this means that the result of Grohe, Kreutzer and
Siebertz~\cite{gks14} captures {\em all} subgraph-closed sparse graph classes on which FO
model checking is fixed parameter tractable.

However, for other types of structures, such as dense graphs or algebraic
structures, the parameterized complexity of FO model checking is largely
uncharted territory.  Grohe~\cite{gro07} notes that ``{\em it would also be
very interesting to study the complexity of model-checking problems on
finite algebraic structures such as groups, rings, fields, lattices, et
cetera}''.  From this perspective it is particularly interesting to
investigate model checking problems on {\em partially ordered sets}
(posets), since posets can be seen both as dense graphs and as algebraic
structures.  Motivated by Grohe's survey~\cite{gro07}, Bova, Ganian and
Szeider~\cite{bgs14, bgs14-ipec} initiated the study of FO model checking on
posets.  As a preliminary result, they show that FO model checking on posets
parameterized by $\phi$ is not fixed parameter tractable unless FPT = W[1],
motivating the study of FO model checking on restricted classes of posets. 
Bova, Ganian and Szeider \cite{bgs14} identified posets of bounded width as a particularly
interesting class to investigate.  Their main technical contribution has been an
FPT algorithm for model checking {\em existential} FO logic on posets of
bounded width, and they left the existence of an FPT algorithm for model
checking FO logic on posets of bounded width as an open problem.  In
subsequent work Gajarsk\'y et al.~\cite{ghoo14} gave a simpler and faster
algorithm for model checking existential FO logic on posets of bounded
width.  Nevertheless, the existence of an FPT algorithm for model checking
full FO logic remained open.

\vspace*{-1ex}
\paragraph{Our contribution.}
In this paper we resolve the open problem of Bova, Ganian and Szeider by
designing a new algorithm for model checking FO logic on posets.
The running time of our algorithm on an $n$-element poset of width $w$ is
$f(\phi,w)\cdot n^2$. Thus our algorithm is not only FPT when parameterized by $\phi$ on posets of
bounded width, it is also FPT by the compound parameter $\phi + w$. 
We demonstrate the generality and
applicability of our main result by showing that a simple
FO-interpretation can be used to obtain an FPT-algorithm for
another natural dense graph class, namely $k$-fold proper interval graphs. 
This generalizes and simplifies the main
result of Ganian et al.~\cite{GHKOST13}.

Our algorithm is based on a new locality lemma for posets. More concretely,
we show that for every poset ${\cal P}$ and formula $\phi$ one can
efficiently iteratively construct a directed graph $D$ such that (a) the vertex set of
$D$ are the elements of ${\cal P}$, (b) every element of ${\cal P}$ has
bounded out-degree in $D$, and (c) it is possible to determine whether
${\cal P} \models \phi$ by checking whether $\phi$ holds on sub-posets of
${\cal P}$ induced by constant-radius balls in $D$.

The statement of our lemma sounds very similar to that of Gaifman's locality
theorem, the crucial differences being that the digraph $D$ is not the Gaifman
graph of ${\cal P}$ and that $D$ depends on the quantifier rank of~$\phi$.  
Indeed, constant radius balls in the Gaifman graph of
constant width posets typically contain the entire poset.  Thus a naive
application of Gaifman's theorem would reduce the problem of deciding
whether ${\cal P}$ is a model of $\phi$ to itself.  The crucial difficulty
we have to overcome is that we have to make the digraph $D$ ``dense enough''
so that (c) holds, while keeping it ``sparse enough'' so that the
vertices in $D$ still have bounded out-degree.  The latter is
necessary to ensure that constant radius balls in $D$ have constant size,
making it feasible to use the naive model checking algorithm for determining
whether $\phi$ holds on sub-posets of ${\cal P}$ induced by constant-radius
balls in $D$.  The construction of the graph $D$ and the proof that it
indeed has the desired properties relies on a delicate inductive argument
thoroughly exploiting properties of posets of bounded width.

\vspace*{-1ex}
\paragraph{Organization of the paper.}In Section~\ref{sec:prelim} we set up
the definitions and the necessary notation. In Section~\ref{sec:poset-types}
we define the digraph $D$ used in our poset locality lemma, and prove some useful structural properties of $D$. In Section~\ref{sec:FOmodelcheck} we prove the locality lemma for posets and give the FPT algorithm for FO logic model checking. We then proceed to show in Section~\ref{sec:interval} how our algorithm can be used to give an FPT algorithm for model checking FO logic on $k$-fold proper interval graphs. 
Finally, in Section~\ref{sec:concl} we conclude with a discussion of further research directions.

\section{Preliminaries}
\label{sec:prelim}

\subsection{Graphs and Posets}
\newcommand{\dist}{\mbox{dist}}

We deal with directed graphs (shortly {\em digraphs}) whose vertices and
arcs bear auxiliary labels, and which may contain parallel arcs.
For a directed graph $D$, a vertex $v \in V(D)$, and an integer
$r$, we denote by $R_{r}^D(v)$, the set of vertices of $D$ that are
reachable from $v$ via a directed path of length at most $r$. Slightly
abusing the notation, we extend the function $R_r^D$ to multiple vertices as
follows; $R_r^D(v_1,\dots,v_k)=\bigcup_{i=1}^k R_{r}^D(v_i)$.
Moreover, for
$v,v'\in V(D)$ we denote by $\dist_D(v,v')$ the length of a shortest directed path from
$v$ to $v'$ in $D$.

A \emph{poset} $\PPP$ is a pair $(P,\leq^\PPP)$ where $P$ is a finite set and $\leq^\PPP$ is
a reflexive, anti-symmetric, and transitive binary relation over $P$.
The \emph{size} of a poset $\PPP=(P,\leq^\PPP)$ is $\Vert\PPP\Vert:=|P|$. 
We say that $p$ and $p'$ are \emph{incomparable}
(in $\PPP$), denoted $p \incom^\PPP\! p'$, if neither $p \leq^\PPP\! p'$ nor $p'
\leq^\PPP\! p$ hold. 
We say that $p'$ is {\em above} $p$
(and $p$ is {\em below}~$p'$) if $p\leq^\PPP\!p'$ and $p\neq p'$.
A \emph{chain} $C$ of
$\PPP$ is a subset of $P$ such that $x \leq^\PPP\! y$ or $y \leq^\PPP\! x$
for every $x,y \in C$. 
A \emph{chain partition} of $\PPP$ is a tuple $(C_1,\dotsc,C_k)$ such
that $\{C_1,\dotsc,C_k\}$ is a partition of $P$ and for every $i$ with
$1 \leq i \leq k$ the poset induced by $C_i$ is a chain of $\PPP$.
An \emph{anti-chain} $A$ of $\PPP$ is a subset
of $P$ such that for all $x,y \in P$ it is true that  $x \incom^\PPP\! y$.
The \emph{width} of a poset $\PPP$, denoted by $\width(\PPP)$ is the 
maximum cardinality of any anti-chain of $\PPP$.
\begin{PRO}[{\cite[Theorem 1.]{frs03}}]\label{pro:comp-chain-part}
  Let $\PPP$ be a poset. Then in time
  \mbox{$\bigO(\width(\PPP)\cdot\Vert\PPP\Vert^2)$}, it is
  possible to compute both $\width(\PPP)=w$ and a
  corresponding chain partition $(C_1,\dotsc,C_{w})$ of $\PPP$.
\end{PRO}

\subsection{Parameterized Complexity}

Here we introduce the most basic concepts of parameterized complexity theory.
For more details, we refer to the many existing text books on the topic~\cite{df99,fg06,nie06}.
An instance of a parameterized problem is a pair $\tuple{x,k}$ where $x$ is
the input and $k$ a parameter.  A~parameterized 
problem $\cal P$ is \emph{fixed-parameter tractable (FPT)} if, for every instance
$\tuple{x,k}$, it can be decided whether $\tuple{x,k}\in\cal P$ 
in time $f(k)\cdot\Card{x}^c$, where $f$ is a
computable function, and $c$ is a constant.

\subsection{First-order Logic}

In this paper we deal with the, well known, relational first-order (FO)
logic. Formulas of this logic are built from (a finite set of) variables,
relational symbols, logical connectives ($\land, \lor, \neg$) and
quantifiers ($\exists, \forall$). A {\em sentence} is a formula with no free
variables. We restrict ourselves to formulas that are in \emph{negation normal
form}; a first-order formula is in negation normal form if all negation symbols
occur only in front of the atoms. Obviously, any first-order formula can
be, in a linear time, converted into an equivalent one in negation normal form. 

The problem we are interested in is a \emph{model checking problem} for
FO formulas on posets, which is formally defined as follows:

\defparproblem{{\sc Poset FO Model Checking}}{A first-order sentence $\phi$ and a poset $\PPP=(P,\leq^\PPP)$.}
{$\width(\PPP)$, $|\phi|$}{Is it true $\PPP\models\phi$, i.e., is $\PPP$ a model of $\phi$?}

\smallskip\noindent
All first-order formulas in this paper are evaluated over posets as follows.
The vocabulary consists of the one binary relation $\leq^\PPP$ and
a finite set of arbitrary unary relations (``colors'' of poset elements).
Atoms of these FO formulas can be equalities between variables ($x=y$),
applications of the predicate $\leq^\PPP$ (with the natural meaning of
$x\leq^\PPP\! y$ in the poset $\PPP$), or applications of one of the unary
predicates $c(x)$ (with the meaning that $x$ is of color~$c$).
For a more detailed treatment of the employed setting, we refer the reader
to~\cite{bgs14}.

While for any fixed sentence $\phi$ one can easily decide whether
$\PPP\models\phi$ in polynomial time, by a brute-force expansion of all the
quantifiers, such a solution is not FPT since the exponent depends on
$\phi$.
Our aim is to provide an FPT solution in the case when we additionally
parameterize the input by the width of~$\PPP$.

It is well known that the model checking problem for almost any logic can be
formulated as finding a winning strategy in an appropriate model checking
game, often called \emph{Hintikka game} (see e.g. \cite{Gradeletal05}). 
In our case the game $\cG(\PPP, \phi)$ for a poset
$\PPP$ and an FO formula $\phi=\phi(x_1,\ldots, x_k)$ in negation normal form (where
$x_1,\ldots, x_k$ are the free variables of $\phi$) is defined as follows:

The game is played by two players, the existential player (Player $\exists$, Verifier), who tries to prove that $\PPP\models \phi$, and the universal
player (Player $\forall$, Falsifier), who tries to disprove that
claim. The positions of this game $\cG(\PPP, \phi)$ are pairs $(\psi,
\beta)$, where $\psi\equiv\psi(x_1,\ldots,x_\ell)$ is a subformula of $\phi$, and
$\beta: \{x_1,\ldots,x_\ell\}\to P$ assigns free variables of $\psi$ elements
of the poset $\PPP$. We write $\psi(p_1,\ldots,p_\ell)$ for a position
$(\psi,\beta)$, where $\beta(x_i)=p_i$ for the free variables $x_i$ of
$\psi$. The initial position of the game is $(\phi, \beta_0)$, where
$\beta_0$ is the initial assignment (if $\phi$ has free variables, or
$\beta_0$ is empty).

The game is played as follows: the existential player (Verifier) moves from
positions associated with disjunctions and formulas starting with the
existential quantifier. From a position $\psi_1\lor\psi_2$ he moves
to either $\psi_1$ or $\psi_2$. From a position 
$\psi(p_1,\ldots,p_i)\equiv \exists y.\, \psi'(p_1,\ldots,p_i,y)$
he moves to any position $\psi'(p_1,\ldots,p_i,p)$, where $p\in P$. The
universal player plays similarly from conjunctions and universally
quantified formulas. At atoms which, in our case, are the positions
$\sigma(p_1)$ of the form $c(p_1)$, or $\sigma(p_1,p_2)$ of
the form $p_1=p_2$, $\neg(p_1=p_2)$, $p_1\leq^\PPP\! p_2$  or
$\neg(p_1\leq^\PPP\! p_2)$, the existential
player wins if $\PPP\models \sigma(p_1,p_2)$, and otherwise the universal
player wins. The equivalence of these games to the standard semantic of FO
is given by the following claim:

\begin{PRO}\label{prop:Hintikka}
  The existential player has a winning strategy in the Hintikka game $\cG(\PPP,
  \phi)$ for a poset $\PPP$ and a first-order sentence $\phi$ if, and only if, $\PPP
  \models \phi$.
\end{PRO}

\section{Poset Structure and Types}
\label{sec:poset-types}

For the rest of the paper we fix a poset $\PPP=(P,\leq^\PPP)$ of width $w$,
a mapping $\lambda:P\to\Lambda$ where $\Lambda$ is a (fixed) finite set of
{\em colors}, and a chain partition $(C_1,\dotsc,C_w)$ of $\PPP$. 
To emphasize that $\PPP$ is associated with an auxiliary coloring~$\lambda$,
we sometimes call $\PPP$ a {\em colored poset}.
For $p \in C_j$ with $1 \leq j \leq w$, we denote by $C(p)$ the chain $C_j$. 
The purpose of this section is to find a description of the
poset $\PPP$ structure suitable for applying ``locality tools'' of finite model theory.
This turns out to be a delicate job requiring a careful inductive definition.

For an integer $s\geq0$, we set $r_s:=3\cdot 4^s-1$ and inductively define
a) a labeling function $\tau_{s} : P \rightarrow \naturals$, and
b) a vertex-labeled and arc-labeled directed graph $D_{s}$
on the vertex set $V(D_s):=P$ as follows:

\begin{DEF}\label{def:graphDs}\rm
For an integer $s\geq0$ and an element $p \in P$, 
we shortly denote by $P_s(p)$ the set $R_{r_{s}}^{D_{s}}(p)$
(i.e., the set of vertices reachable in $D_{s}$ from $p$ at distance $\leq r_s$). 
We let $\tau_{0}(p):=\langle \lambda(p),j\rangle$, where $j$ is the index s.t.~$C_j=C(p)$.
Inductively for every integer $s\geq0$, we define
$D_{s}$ as the digraph with the vertex set $P$ and vertex labels
given by $\tau_{s}$, containing the following arcs: 
\begin{itemize}\parskip-1pt
\item for every $p \in P$ and every $j\in\{1,\dots,w\}$,
  $D_{s}$ contains an arc from $p$ with label $\elmax$
  to the topmost element of $C_j$;
\item for every $p \in P$ and every $j\in\{1,\dots,w\}$,
  $D_{s}$ contains an arc from $p$ with label $\elmin$
  to the bottommost element of $C_j$;
\item for every $p \in P$, every $j\in\{1,\dots,w\}$, and
  every $t \in \SB \tau_{s}(q) \SM q \in C_j \SE$,
  $D_{s}$ contains an arc from $p$ to $p'$ with label $\elup$, where $p'\not=p$ is the
  bottommost element of $C_j$ such that $\tau_{s}(p')=t$ and $p \leq^\PPP\!
  p'$ (if such an element $p'$ exists);
\item for every $p \in P$, every $j\in\{1,\dots,w\}$, and
  every $t \in \SB \tau_{s}(q) \SM q \in C_j \SE$, $D_{s}$
  contains an arc from $p$ to $p'$ with label $\eldown$, where $p'\not=p$ is the
  topmost element of $C_j$ such that $\tau_{s}(p')=t$ and $p' \leq^\PPP\!
  p$ (if such an element $p'$ exists).
\end{itemize}
Then, having defined $D_s$ as above, we set for every element $p \in P$
\begin{itemize}\parskip-1pt\item
$\tau_{s+1}(p)$ to be the isomorphism type of a relational structure
$\AAA_s^\PPP(p)$, where $\AAA_s^\PPP(p)$ is formed by the vertex- and arc-labeled
induced subdigraph $D_{s}[P_{s}(p)]$ rooted at~$p$ with the additional
binary relation $\leq^\PPP$ restricted to~$P_{s}(p)$.
\end{itemize}
\end{DEF}

The values $\tau_{s}(p)$, $p\in P$, will also be called the {\em types of rank~$s$}
(of elements of $\PPP$), where the rank will often be implicit from the context.
It is useful to notice that the considered coloring $\lambda$ of the poset
$\PPP$ elements is fully determined by their types of rank~$0$ in $D_0$
(and so also by their types of any higher rank).
Therefore, we may skip an explicit reference to $\lambda$ in the rest of this
section.

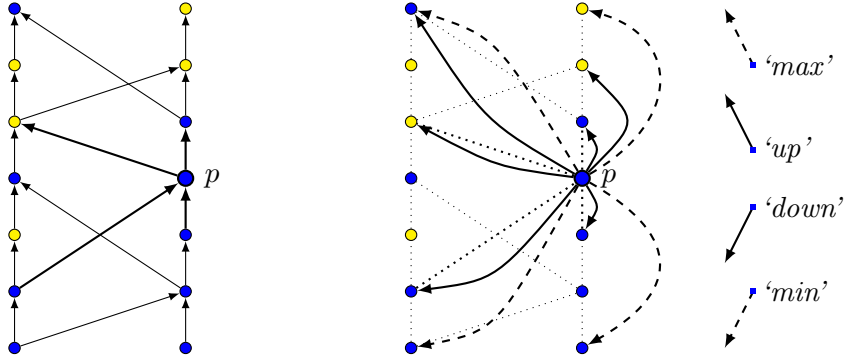
\begin{figure}[tb]
$$
\begin{tikzpicture}[scale=0.75]
\tikzstyle{every node}=[draw, shape=circle, minimum size=2pt,inner sep=1.5pt, fill=blue]
\draw (0,0) node (a) {}; \draw (0,1) node (b) {};
\draw (0,2) node[fill=yellow] (c) {}; \draw (0,3) node (d) {};
\draw (0,4) node[fill=yellow] (e) {}; \draw (0,5) node[fill=yellow] (f) {};
\draw (0,6) node (g) {};
\draw (3,0) node (A) {}; \draw (3,1) node (B) {};
\draw (3,2) node (C) {};
\draw (3,3) node[thick,inner sep=2pt, label=right:$p$] (D) {};
\draw (3,4) node (E) {}; \draw (3,5) node[fill=yellow] (F) {};
\draw (3,6) node[fill=yellow] (G) {};
\tikzstyle{every path}=[->]
\draw (a) -- (b); \draw (b) -- (c); \draw (c) -- (d);
\draw (d) -- (e); \draw (e) -- (f); \draw (f) -- (g);
\draw (A) -- (B); \draw (B) -- (C); \draw[thick] (C) -- (D);
\draw[thick] (D) -- (E); \draw (E) -- (F); \draw (F) -- (G);
\draw[thick] (b) -- (D);
\draw[thick] (D) -- (e);
\draw (a) -- (B); \draw (B) -- (d);
\draw (E) -- (g); \draw (e) -- (F);
\end{tikzpicture}
\qquad\qquad\qquad
\begin{tikzpicture}[scale=0.75]
\tikzstyle{every node}=[draw, shape=circle, minimum size=2pt,inner sep=1.5pt, fill=blue]
\draw (0,0) node (a) {}; \draw (0,1) node (b) {};
\draw (0,2) node[fill=yellow] (c) {}; \draw (0,3) node (d) {};
\draw (0,4) node[fill=yellow] (e) {}; \draw (0,5) node[fill=yellow] (f) {};
\draw (0,6) node (g) {};
\draw (3,0) node (A) {}; \draw (3,1) node (B) {};
\draw (3,2) node (C) {}; 
\draw (3,3) node[thick,inner sep=2pt, label=right:$p$] (D) {};
\draw (3,4) node (E) {}; \draw (3,5) node[fill=yellow] (F) {};
\draw (3,6) node[fill=yellow] (G) {};
\tikzstyle{every path}=[dotted]
\draw (a) -- (b); \draw (b) -- (c); \draw (c) -- (d);
\draw (d) -- (e); \draw (e) -- (f); \draw (f) -- (g);
\draw (A) -- (B); \draw (B) -- (C); \draw[thick] (C) -- (D);
\draw[thick] (D) -- (E); \draw (E) -- (F); \draw (F) -- (G);
\draw[thick] (b) -- (D);
\draw[thick] (D) -- (e);
\draw (a) -- (B); \draw (B) -- (d);
\draw (E) -- (g); \draw (e) -- (F);
\tikzstyle{every path}=[solid,thick,->]
\draw[dashed] (D) .. controls (1.5,0.3) .. (a);
\draw[dashed] (D) .. controls (1.5,5.5) .. (g);
\draw[dashed] (D) .. controls (4.75,4) and (4.75,5.5) .. (G);
\draw[dashed] (D) .. controls (4.75,2) and (4.75,1) .. (A);
\draw (D) .. controls (1.5,1.2) .. (b);
\draw (D) .. controls (1.5,3.2) .. (e);
\draw (D) .. controls (1.3,4) .. (g);
\draw (D) .. controls (3.3,3.5) .. (E);
\draw (D) .. controls (3.3,2.5) .. (C);
\draw (D) .. controls (4,4) .. (F);
\tikzstyle{every node}=[inner sep=1pt, fill=blue]
\draw[dashed] (6,1) node[label=right:$\elmin$] {} -- (5.5,0);
\draw[dashed] (6,5) node[label=right:$\elmax$] {} -- (5.5,6);
\draw (6,3.5) node[label=right:$\elup$] {} -- (5.5,4.5);
\draw (6,2.5) node[label=right:$\eldown$] {} -- (5.5,1.5);
\end{tikzpicture}
$$
\caption{The picture, on the left, shows an upward-directed
Hasse diagram of a bicolored poset $\PPP$
(where $\PPP$ is the reflexive and transitive closure of it).
On the right, the picture shows the arcs of $D_0$ starting from a selected
element~$p\in P$, as by Definition~\ref{def:graphDs}.}
\label{fig:D0}
\end{figure}

The definition of $D_0$ is illustrated in Figure~\ref{fig:D0}.
Informally, the type of an element $p\in P$ captures its ``local neighborhood''
(which is growing in size with the rank~$s$), and
the digraph $D_s$ contains $\elup$-arcs ($\eldown$-arcs)
from $p$ to the next higher (next lower) elements of $\PPP$ of each appearing type.
Moreover, there are shortcut arcs, labeled $\elmin$ and $\elmax$, from $p$ to
the extreme elements of each chain of $\PPP$.
It is important that, since we use a fixed finite number of colors in $\PPP$
and since $\PPP$ is of bounded width, the outdegrees in $D_s$ are
inductively bounded for every fixed~$s$ independently of the size of~$\PPP$.

For start we need the following basic properties of the digraph~$D_s$
and the labeling function $\tau_s$, which are easy to prove.
The first two of these simple claims establish that the sequence of labeled digraphs
$D_0,D_1,D_2,\dots$ indeed presents an increasingly finer resolution of a ``local
structure'' of the poset~$\PPP$.
For all the claims, let $\PPP=(P,\leq^\PPP)$ be a poset and 
$P_s,D_s,\AAA_s^\PPP$ and $\tau_s$ be as in Definition~\ref{def:graphDs}.

\begin{LEM}\label{lem:coarsening-types}
  For every $p, p' \in P$ and $s\geq0$, if $\tau_{s}(p) \neq \tau_{s}(p')$,
  then also $\tau_{s+1}(p) \neq \tau_{s+1}(p')$.
\end{LEM}
\begin{proof}
  Since $\tau_{s}(p) \neq \tau_{s}(p')$, it follows that $p$ and
  and $p'$ have different vertex labels in $D_{s}$. Hence, the
  isomorphism types of $\AAA_s^\PPP(p)$ 
  and
  $\AAA^\PPP_s(p')$ 
  are not the same either and hence
  $\tau_{s+1}(p) \neq \tau_{s+1}(p')$.
\end{proof}

\begin{LEM}\label{lem:dgraph-edge-preservation}
  For every $s\geq0$, if $D_{s}$ contains an arc from
  some vertex $p \in P$ to some $p' \in P$, then $D_{s+1}$
  also contains an arc from $p$ to $p'$ with the same label.
  In other words, $D_{s}$ is a spanning subdigraph of $D_{s+1}$ (neglecting
  the vertex-labels).
\end{LEM}
\begin{proof}
  Let $p, p' \in P$ and assume there is an arc $a$ from $p$ to $p'$
  in $D_{s}$. We need to show that there is also an arc from
  $p$ to $p'$ in $D_{s+1}$ that has the same label as~$a$. Depending
  on the arc-label of $a$ in $D_{s}$ we distinguish the following cases:
  \begin{enumerate}
  \item Assume that the label of $a$ is $\elup$. Then, $p'$ is the
    bottommost element of type $\tau_{s}(p')$ on the chain $C(p')$ such
    that $p \lneq^P\! p'$. Hence, for any element $p''\in C(p')$ 
    such that $p \lneq^P\! p'' \lneq^P\! p'$, it holds that $\tau_{s}(p'') \neq
    \tau_{s}(p')$. By Lemma~\ref{lem:coarsening-types}, we
    obtain that also $\tau_{s+1}(p'') \neq \tau_{s+1}(p')$. Consequently, $p'$ is
    also the bottommost element of type $\tau_{s+1}(p')$ on the chain $C(p')$
    with $p \lneq^P\! p'$. This shows that $D_{s+1}$ contains an arc
    from $p$ to $p'$ with label $\elup$, as required.
  \item The argument for the case of label $\eldown$ is
    similar to the previous case.
  \item The claim trivially holds for the labels $\elmin$ and $\elmax$.
  \vspace*{-3ex}
  \end{enumerate}
\end{proof}

Another simple property of Definition~\ref{def:graphDs}
is that pairs of arcs of the same vertex- and arc-labels in the digraph $D_s$
never ``cross one another'', which is formalized as follows:

\begin{LEM}\label{lem:nocrossing-Ds}
  Let $\PPP=(P,\leq^\PPP)$ be a poset and $P_s,D_s$ and $\tau_s$ be as in
  Definition~\ref{def:graphDs}. Assume that $p,p',q,q'\in P$ are such that $\tau_{s}(p)=\tau_{s}(p')$
  and $\tau_{s}(q)=\tau_{s}(q')$,
  and that both $(p,q)$ and $(p',q')$ are arcs of the same label in $D_s$.
  If $p \leq^\PPP\! p'$ then $q \leq^\PPP\! q'$.
\end{LEM}
\begin{proof}
  Note that $p'\in C(p)$ and $q'\in C(q)$, and so the pairs are comparable.
  If the label of $(p,q)$ is $\elmin$ or $\elmax$, then $q=q'$ and the claim holds.
  Assume that the label of $(p,q)$ is $\elup$ and that $q' \lneq^P\! q$.
  Then $p \leq^\PPP\! p' \leq^\PPP\! q'$ and so, by Definition~\ref{def:graphDs},
  the arc $(p,q)$ should point from $p$ to $q'$ in $D_s$, a contradiction.
  Therefore, $q \leq^\PPP\! q'$ as required.
  The case of label $\eldown$ is similar.
\end{proof}

The subsequent claims are more involved and technical.
Informally, they together show that for any $p_1,\dotsc,p_k \in P$,
a property or relation of other element(s) of $\PPP$ to $p_1,\dotsc,p_k$ can
also be observed in a given bounded neighborhood of $p_1,\dotsc,p_k$ in~$D_s$.
Importantly, the richer local property is observed, the higher index
$s$ in $D_s$ is used.
The easy base case of $s=0$ is covered by
Lemma~\ref{lem:equivalent-element-in-neighborhood}
while the general case of $s$ is inductively established by
Lemma~\ref{lem:in-neighborhood} and reformulated in
Corollary~~\ref{cor:isomorphic-neighborhoods}.
We refer to Section~\ref{sec:FOmodelcheck} for details on using these claims.

\begin{LEM}\label{lem:equivalent-element-in-neighborhood}
  Let $\PPP=(P,\leq^\PPP)$ be a poset and $D_0$ the digraph defined in
  Definition~\ref{def:graphDs}.
  For any $k\geq1$ and $p,\, p_1,\dotsc,p_k \in P$, there exists an element
  $p' \in R_{2}^{D_{0}}(p_1,\dotsc,p_k)$
  such that $\tau_{0}(p')=\tau_{0}(p)$, and $p',p$ are in the same relation with
  respect to all of $p_1,\dotsc,p_k$ in $\PPP$:
  formally, for every $i\in\{1,\dots,k\}$, it holds that
  $p' \leq^\PPP\! p_i$ if and only if $p \leq^\PPP\! p_i$, and 
  $p_i \leq^\PPP\! p'$ if and only if $p_i \leq^\PPP\! p$.
\end{LEM}
\begin{proof}
  Choose $p'$ to be the topmost element of the chain $C(p)$ such that
  $\tau_{0}(p')=\tau_{0}(p)$ and,
  for every element $q\in C(p)$ with $p \lneq^\PPP\! q \leq^\PPP\! p'$,
  it holds that $q \notin R_1^{D_{0}}(p_1,\dotsc,p_k)$.
  Then, clearly, $p\leq^\PPP\! p'$ (while it might happen $p=p'$)
  and $p' \in R_{2}^{D_{0}}(p_1,\dotsc,p_k)$ by the definition.
  It remains to show that $p'$ is in the same relation with respect to
  $p_1,\dotsc,p_k$ in $\PPP$ as $p$. Assume for a contradiction that
  this is not the case. Then there is an index $i\in\{1,\dots,k\}$
  such that $p$ and $p'$ are in a different relation towards $p_i$.
  Because $p \leq^\PPP\! p'$, we have that either (I) $p \leq^\PPP\! p_i$ but
  $p' \not\leq^\PPP\! p_i$, or (II) $p_i \leq^\PPP\! p'$ but $p_i \not\leq^\PPP\! p$.

  In the case (I), either $p_i \in C(p)$ and hence $p_i$ is in between 
  $p$ and $p'$ on the chain $C(p)$ contradicting our choice of $p'$, or
  $p_i \notin C(p)$ but $D_{0}$ contains an arc with label $\eldown$ from
  $p_i$ to some element on the chain $C(p)$ between $p$ and $p'$, which
  again contradicts our choice of $p'$. 
  In the case (II), either $p_i \in C(p)$ and hence $p_i$ is in between 
  $p$ and $p'$ on the chain $C(p)$ contradicting our choice of $p'$, or
  $p_i \notin C(p)$ but $D_{0}$ contains an arc with label $\elup$ from
  $p_i$ to some element on the chain $C(p)$ between $p$ and $p'$, which
  again contradicts our choice of $p'$. 
\end{proof}

We use the following shorthand notation.
For $p,p',q\in P$ we say that $q$ {\em discerns $p$ from $p'$}, with respect
to $\leq^\PPP$, if (at least) one of the following four conditions holds true;
$p \leq^\PPP\! q$ and $p' \not\leq^\PPP\! q$,~
$p \not\leq^\PPP\! q$ and $p' \leq^\PPP\! q$,~
$q \leq^\PPP\! p$ and $q \not\leq^\PPP\! p'$, or
$q \not\leq^\PPP\! p$ and $q \leq^\PPP\! p'$.
For example, the conclusion of
Lemma~\ref{lem:equivalent-element-in-neighborhood} is equivalent to saying
``neither of $p_1,\dotsc,p_k$ discerns $p$ from $p'$''.
Since the poset $\PPP$ is fixed for this section, we will often skip an
explicit reference to~$\leq^\PPP$.

\begin{LEM}\label{lem:in-neighborhood}
  Let $\PPP=(P,\leq^\PPP)$ be a poset, $s\geq0$ an integer,
  $D_{s}$ and $\tau_s$ be as in Definition~\ref{def:graphDs}, and $p, p' \in P$.
  Assume that $p \leq^\PPP\! p'$ and $\tau_{s+1}(p)=\tau_{s+1}(p')$,
  where the latter is witnessed by an isomorphism
  $\iota: P_{s}(p) \to P_{s}(p')$
  of the structures $\AAA^{\PPP}_{s}(p)$ and $\AAA^{\PPP}_{s}(p')$.
  If $a \in P$ and $b \in P_{s}(p)$ are such that $a$ discerns
  $b$ from $\iota(b)$ (with respect to $\leq^\PPP$), then
  there exists a directed path from $a$ to some element $a' \in C(p)$
  of type $\tau_{s+1}(p)$ with $p \leq^\PPP\! a' \leq^\PPP\! p'$, in $D_{s}$
  of length at most $2\ell+1$ where $\ell=\dist_{D_{s}}(p,b)$.
\end{LEM}

\begin{figure}[tb]
$$
\begin{tikzpicture}[scale=0.75]
\draw[thick] (2.1,1) ellipse (30mm and 12mm);
\draw[thick] (2.1,6) ellipse (30mm and 12mm);
\draw[color=gray] (2,3.4) ellipse (6mm and 35mm);
\draw (3,4) node[label=right:$\leq2\ell+1$] {};
\draw (-0.75,1) node[label=left:$P_{s}(p)$] {};
\draw (-0.75,6) node[label=left:$P_{s}(p')$] {};
\draw (-1,3.2) node[label=left:$\iota\!\!$] (iota) {};
\draw[color=gray,line width=2.5mm,->] (0,2.1) .. controls (iota) .. (0,5);
\tikzstyle{every node}=[draw, shape=circle, minimum size=2pt,inner sep=1.5pt,fill=yellow]
\draw (2,1) node[fill=blue, label=left:$p$~~~] (p) {}; 
\draw (2,6) node[fill=blue, label=left:{$p'=\iota(p)$~~~}] (pp) {};
\draw (7,3.6) node[label=right:$a$] (a) {};
\draw (2,3.6) node[fill=blue, label=left:$a'$~~~~] (aa) {};
\draw (4,1) node[label=right:$b$, label=left:$\ell$~~~~] (b) {}; 
\draw (4,6) node[label=right:$\iota(b)$, label=left:$\ell$~~~~] (bb) {}; 
\draw (2,2.4) node {}; \draw (2,2.9) node {};
\draw (2,4.1) node {}; \draw (2,4.5) node {};
\tikzstyle{every path}=[->]
\draw[color=gray,dashed] (p) -- (aa) ;
\draw[color=gray,dashed] (aa) -- (pp);
\draw[color=gray,dashed,thick] (b) -- (a);
\draw[color=gray,dashed,thick] (a) -- (bb);
\draw (p) -- (2.8,0.7) -- (3.3,1.5) -- (b);
\draw (pp) -- (2.8,5.7) -- (3.3,6.5) -- (bb);
\draw[thick] (a) -- (6,3.3) -- (5,3.8) -- (4,3) -- (3,4) -- (aa);
\end{tikzpicture}
$$
\caption{An illustration of the statement of Lemma~\ref{lem:in-neighborhood}:
the dashed arcs depict the poset relation $\leq^\PPP$ while the solid arcs
represent directed paths in the digraph $D_s$.
}
\label{fig:in-neighborhood}
\end{figure}
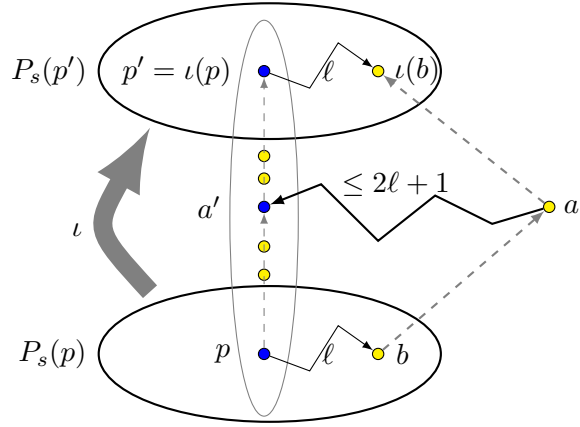

The technical statement of Lemma~\ref{lem:in-neighborhood}
deserves an informal explanation.
For start, if elements $p,p'\in P$ such that $p'\in C(p)$ are discerned in the poset
$\PPP$ by an element $a\in P$
then, clearly, $a=a'\in C(p)$ or the Hasse diagram of $\PPP$ contains an arc
between $a$ and some $a'\in C(p)$ such that $a'$ lies between $p$ and $p'$
on~$C(p)$.
Lemma~\ref{lem:in-neighborhood} then largely extends this simple observation
to the setting of Definition~\ref{def:graphDs} and for discerned elements in
neighborhoods of $p$ and~$p'$.
The statement is illustrated in Figure~\ref{fig:in-neighborhood}.

\begin{proof}[Proof of Lemma~\ref{lem:in-neighborhood}]
  Notice that the assumption $\tau_{s+1}(p)=\tau_{s+1}(p')$ immediately implies
  $p'\in C(p)$.
  Let $F=(b_0=p,$ $b_1,\dotsc,b_\ell=b)$ be a shortest directed path from $p$ to
  $b$ in $D_{s}$. Then,
  $\iota(F)=\big(\iota(b_0)=p',\iota(b_1),\dotsc,\iota(b_\ell)=\iota(b)\big)$
  is a directed path in $D_{s}$ that has the same arc-labels and
  vertex-labels as $F$.   
  Observe that since $a$ discerns $b$ from $\iota(b)$,
  it holds that $b \neq \iota(b)$ and hence also $b_i \neq
  \iota(b_i)$ for every $i\in\{0,\dots,\ell\}$.
  Moreover, $p \leq^\PPP\! p'=\iota(p)$ implies $b \leq^\PPP\! \iota(b)$
  by a simple inductive argument using Lemma~\ref{lem:nocrossing-Ds} along
  the path~$F$.

  We will show the existence of $a' \in C(p)$ and of the required path from $a$ to $a'$ 
  in $D_{s}$ by proving the following slightly stronger statement:
\begin{itemize}\item
  For every $i\in\{0,\dots,\ell\}$, there is an
  element $a_i \in C(b_i)$ of type $\tau_{s+1}(b_i)$ 
  such that $b_i \leq^\PPP\! a_i \leq^\PPP\! \iota(b_i)$ and $D_{s}$ contains
  a directed path from $a$ to $a_i$ of length at most $2(\ell-i)+1$. 
\end{itemize}
  We prove the statement via induction on $(\ell-i)$ starting from $i=\ell$.

  For $i=\ell$, we obtain $a_\ell$ as follows. If $a \in C(b)$, 
  then $b \leq^\PPP\! a \leq^\PPP\! \iota(b)$ because $a$ discerns $b$ from $\iota(b)$, and
  let $a_\ell \in C(b)$ be the topmost element of type $\tau_{s+1}(b)$ such
  that $a_\ell \leq^\PPP\! a$. Clearly, $b \leq^\PPP\! a_\ell \leq^\PPP\! \iota(b)$ and 
  $D_{s}$ contains an arc with label $\eldown$ from $a$ to
  $a_\ell$, as required.
  So assume that $a \notin C(b)$. Then, again, $b \leq^\PPP\! \iota(b)$ and $a$
  discerns $b$ from $\iota(b)$, and so at least one of
  $b \leq^\PPP\! a$ or $a \leq^\PPP\! \iota(b)$ holds true.

  In the former case, 
  let $a_\ell \in C(b)$ be the topmost element of type $\tau_{s+1}(b)$ such
  that $a_\ell \leq^\PPP\! a$. Since $b \leq^\PPP\! a$ but $\iota(b) \not\leq^\PPP\! a$,
  it follows that $b \leq^\PPP\! a_\ell \leq^\PPP\! \iota(b)$. Furthermore, $D_{s}$
  contains an arc with label $\eldown$ from $a$ to $a_\ell$, as required.
  In the later case, let $a_\ell \in C(b)$ be the bottommost element of
  type $\tau_{s+1}(b)$ such that $a \leq^\PPP\! a_\ell$.
  Since $a \leq^\PPP\! \iota(b)$ but $a \not\leq^\PPP\! b$, it again
  follows that $b \leq^\PPP\! a_\ell \leq^\PPP\! \iota(b)$. Furthermore, $D_{s}$
  contains an arc with label $\elup$ from $a$ to $a_\ell$, as required.

  \smallskip
  Now, consider $i\in\{0,\dots,\ell-1\}$ and that we have already
  shown the claim for $i+1$. Hence, by the induction hypothesis, we can
  assume that there exists $a_{i+1} \in C(b_{i+1})$ of type
  $\tau_{s+1}(b_{i+1})$ with $b_{i+1} \leq^\PPP\! a_{i+1} \leq^\PPP\! \iota(b_{i+1})$
  such that $D_{s}$ contains
  a directed path from $a$ to $a_{i+1}$ of length at most $2(\ell-i-1)+1$.

\begin{figure}[tb]
$$
\begin{tikzpicture}[scale=0.75]
\draw[color=gray] (2,3.4) ellipse (5mm and 32mm);
\draw[color=gray] (5,3.4) ellipse (5mm and 32mm);
\draw[color=gray] (7.2,3.4) ellipse (5mm and 32mm);
\tikzstyle{every node}=[draw, shape=circle, minimum size=2pt,inner sep=1.5pt,fill=yellow]
\draw (2,1) node[fill=blue, label=left:$p$~~~] (p) {}; 
\draw (2,6) node[fill=blue, label=left:{$p'$~~~}] (pp) {};
\draw (11.2,3.6) node[inner sep=2pt, label=right:$a$, 
		label=left:{\scriptsize$\leq2(\ell\!-\!i)\!-\!1$}~] (a) {};
\draw (2,3.6) node[fill=blue, inner sep=2pt,
		label=left:$a'$~~~~, label=right:~~~~\dots] (aa) {};
\draw (5,1) node[fill=blue, label=left:$b_{i}$~~, label=right:~~$\elup$] (b1) {}; 
\draw (5,5) node[fill=blue, label=left:{$\iota(b_{i})$~~~}, label=right:~~$\elup$] (bb1) {};
\draw (5,3.4) node[fill=blue, inner sep=2pt,
		label=left:$a_{i}$~~~~] (a1) {};
\draw (7.2,2) node[fill=brown, 
		label=right:~~$b_{i+1}$~~\dots] (b2) {}; 
\draw (7.2,6) node[fill=brown, 
		label=right:~{$\iota(b_{i+1})$~~\dots}] (bb2) {};
\draw (7.2,4) node[fill=brown, inner sep=2pt,
		label=right:~~~$a_{i+1}$, label=left:{\small$\eldown$}~~~] (a2) {};
\draw (5,2.1) node {}; \draw (5,2.6) node {};
\draw (5,4.1) node {}; \draw (5,4.5) node {};
\draw (7.2,2.4) node {}; \draw (7.2,2.9) node {};
\draw (7.2,4.6) node {}; \draw (7.2,5.1) node {};
\tikzstyle{every path}=[->]
\draw[color=gray,dashed] (p) -- (aa); \draw[color=gray,dashed] (aa) -- (pp);
\draw[color=gray,dashed] (b1) -- (a1); \draw[color=gray,dashed] (a1) -- (bb1);
\draw[color=gray,dashed] (b2) -- (a2); \draw[color=gray,dashed] (a2) -- (bb2);
\draw (p) -- (3.2,0.7) -- (4.3,1.5) -- (b1);
\draw (pp) -- (3.2,5.3) -- (4.3,5.6) -- (bb1);
\draw (b1) -- (b2); \draw (bb1) -- (bb2);
\draw[thick] (a) -- (10,4.6) -- (8,3.1) -- (a2);
\draw[thick] (a2) -- (a1);
\end{tikzpicture}
$$ $$
\begin{tikzpicture}[scale=0.75]
\draw[color=gray] (2,3.4) ellipse (5mm and 32mm);
\draw[color=gray] (5,3.4) ellipse (5mm and 32mm);
\draw[color=gray] (7.2,3.4) ellipse (5mm and 32mm);
\tikzstyle{every node}=[draw, shape=circle, minimum size=2pt,inner sep=1.5pt,fill=yellow]
\draw (2,1) node[fill=blue, label=left:$p$~~~] (p) {}; 
\draw (2,6) node[fill=blue, label=left:{$p'$~~~}] (pp) {};
\draw (11.2,4.5) node[inner sep=2pt, label=right:$a$, 
		label=left:{\scriptsize$\leq2(\ell\!-\!i)\!-\!1$}~~] (a) {};
\draw (2,3.6) node[fill=blue, inner sep=2pt,
		label=left:$a'$~~~~, label=right:~~~~\dots] (aa) {};
\draw (5,1) node[fill=blue, label=left:$b_{i}$~~, label=right:~~$\elup$] (b1) {}; 
\draw (5,5) node[fill=blue, 
		label=left:{$\iota(b_{i})$~~~}, label=right:~~$\elup$] (bb1) {};
\draw (5,3.4) node[fill=blue, inner sep=2pt,
		label=left:$a_{i}$~~~~] (a1) {};
\draw (7.2,2) node[fill=brown, 
		label=right:~~$b_{i+1}$~~\dots] (b2) {}; 
\draw (7.2,6) node[fill=brown, inner sep=2pt,
		label=left:{\scriptsize$\eldown$~~~~~},
		label=right:~~{$\iota(b_{i+1})=a_{i+1}$}] (bb2) {};
\draw (7.2,4) node[fill=brown, label=right:~~~$h$,
		label=left:{\small$\eldown$}~~~] (a2) {};
\draw (5,2.1) node {}; \draw (5,2.6) node {};
\draw (5,4.1) node {}; \draw (5,4.5) node {};
\draw (7.2,2.4) node {}; \draw (7.2,2.9) node {};
\draw (7.2,4.6) node {}; \draw (7.2,5.1) node {};
\draw (5,5.6) node[fill=blue] (bbb1) {};
\tikzstyle{every path}=[->]
\draw[color=gray,dashed] (p) -- (aa); \draw[color=gray,dashed] (aa) -- (pp);
\draw[color=gray,dashed] (b1) -- (a1); \draw[color=gray,dashed] (a1) -- (bb1);
\draw[color=gray,dashed] (b2) -- (a2); \draw[color=gray,dashed] (a2) -- (bb2);
\draw[color=gray,dashed] (bb1) -- (bbb1);
\draw (p) -- (3.2,0.7) -- (4.3,1.5) -- (b1);
\draw (pp) -- (3.2,5.3) -- (4.3,5.6) -- (bb1);
\draw (b1) -- (b2); \draw (bb1) -- (bb2);
\draw[dashed] (bb2) -- (bbb1);
\draw[thick] (a) -- (10,5.1) -- (8.8,4.7) -- (bb2);
\draw[thick] (a2) -- (a1);
\draw[thick] (bb2) .. controls (7.6,5) .. (a2);
\end{tikzpicture}
$$
\caption{An illustration of the proof of Lemma~\ref{lem:in-neighborhood}:
top --
the inductive step in the case of $\iota(b_{i}) \not\leq^\PPP\! a_{i+1}$;
bottom --
the inductive step in the case of $\iota(b_{i}) \leq^\PPP\! a_{i+1}$.
}
\label{fig:proof-in-neighborhood}
\end{figure}
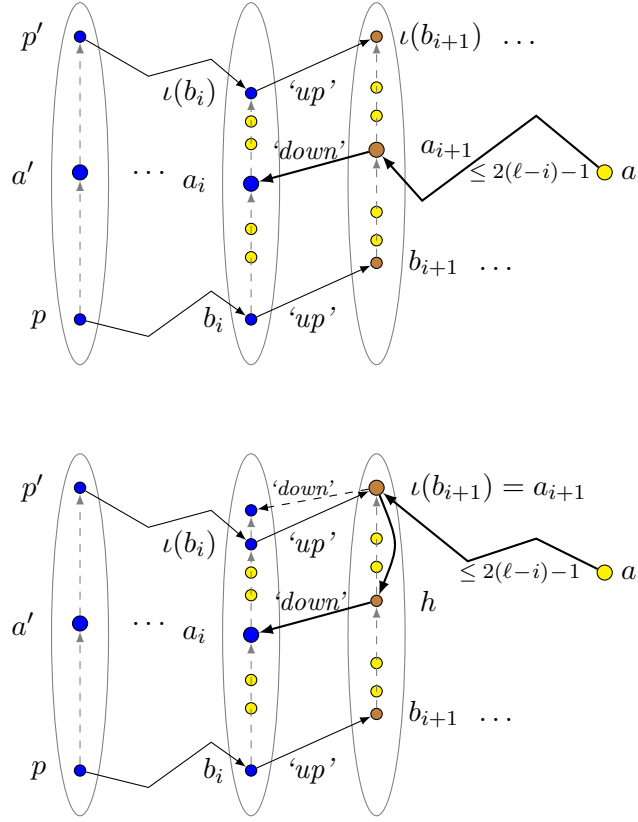

  We need to show that there is $a_i \in C(b_i)$ of type
  $\tau_{s+1}(b_i)$ with $b_i \leq^\PPP\! a_i \leq^\PPP\! \iota(b_i)$ such that
  $D_{s}$ contains a directed path of length at most $2$ from
  $a_{i+1}$ to $a_i$.
  We distinguish the following cases, depending on the arc-label of the
  arc $(b_i,b_{i+1})$ in $D_{s}$.
  \begin{enumerate}\parskip0pt
  \item
    Assume that the label of $(b_i,b_{i+1})$ is $\elup$.
    See Figure~\ref{fig:proof-in-neighborhood}.

    If $\iota(b_{i}) \not\leq^\PPP\! a_{i+1}$, then let $a_i\in C(b_{i})$ be
    the topmost element in $C(b_{i})$ of
    type $\tau_{s+1}(b_i)$ such that $a_i \leq^\PPP\! a_{i+1}$. Then, 
    $D_{s}$ contains an arc with label $\eldown$ from $a_{i+1}$ to $a_i$
    (or, possibly, $a_{i}=a_{i+1}$).
    Since $\iota(b_{i}) \leq^\PPP\! a_{i}$ would contradict $\iota(b_{i})
    \not\leq^\PPP\! a_{i+1}$, we have $a_i \leq^\PPP\! \iota(b_i)$.
    Since $b_i \leq^\PPP\! b_{i+1} \leq^\PPP\! a_{i+1}$ and $a_i$ is the topmost
    one of its kind in $C(b_{i})$, it holds that $b_i \leq^\PPP\! a_i$, as required.

    Otherwise, $\iota(b_{i}) \leq^\PPP\! a_{i+1}$ and because the label of
    the arc $(\iota(b_i),\iota(b_{i+1}))$ in $D_{s}$ is $\elup$, we have
    $a_{i+1}=\iota(b_{i+1})$.
    Let $h$ be the topmost element in $C(b_{i+1})$ of type $\tau_{s+1}(b_{i+1})$
    with $h \lneq^P\! a_{i+1}$.
    Then, $D_{s}$ contains an arc with label $\eldown$ from
    $a_{i+1}$ to $h$ and $b_{i+1} \leq^\PPP\! h$
    (recall $b_{i+1}\not=\iota(b_{i+1})$).
    Because of (the existence of) the arc $(\iota(b_i),\iota(b_{i+1}))$ in $D_{s}$,
    we obtain that $\iota(b_i) \not\leq^\PPP\! h$.
    Let $a_i\in C(b_{i})$ be the topmost element in $C(b_{i})$ of
    type $\tau_{s+1}(b_i)$ such that $a_i \leq^\PPP\! h$.
    Then, $D_{s}$ contains an arc with label $\eldown$ from
    $h$ to $a_i$ and hence a directed path from $a_{i+1}$ to $a_i$ of length~$2$
    (or, possibly, $a_{i}=h$).   
    Clearly, $a_i \leq^\PPP\! \iota(b_i)$.
    Since $b_i \leq^\PPP\! b_{i+1} \leq^\PPP\! h$ and $a_i$ is the topmost
    one of this kind in $C(b_{i})$, it holds that $b_i \leq^\PPP\! a_i$, as required.

  \item 
    Assume that the label of $(b_i,b_{i+1})$ is $\eldown$.
    The situation is completely symmetric to the previous case, and the same
    arguments prove the conclusion (just exchanging $\leq^\PPP$ with
    $\geq^\PPP$ and $\eldown$ with $\elup$).
  \item
    Assume that the label of $(b_i,b_{i+1})$ is $\elmax$ or $\elmin$.
    That would imply $b_{i+1}=\iota(b_{i+1})$, a contradiction.
    This case cannot happen.
  \vspace*{-3ex}
  \end{enumerate}
\end{proof}

The way we shall use Lemma~\ref{lem:in-neighborhood} in
Section~\ref{sec:FOmodelcheck} can be informally summarized as follows.
If $p$ and $p'$ are elements of the same type which are next
to each other on their chain in $\PPP$,
then their $D_s$-neighborhoods appear ``the same'' with respect to
$\leq^\PPP$ to all poset elements which are sufficiently far away from $p$ 
as measured by $D_{s+1}$.
The precise formulation is next.

\begin{corollary}\label{cor:isomorphic-neighborhoods}
  Let $\PPP=(P,\leq^\PPP)$ be a poset, $s\geq0$ an integer,
  the digraphs $D_{s}$ and $D_{s+1}$ be as in Definition~\ref{def:graphDs},
  $p\not=p' \in P$ be such that $\tau_{s+1}(p)=\tau_{s+1}(p')$,
  and the isomorphism map $\iota: P_{s}(p) \to P_{s}(p')$
  be as in Lemma~\ref{lem:in-neighborhood}.
  Assume, moreover, that $D_{s+1}$ contains an arc $(p',p)$ with
  label $\eldown$.
  For any given $p_1,\dotsc,p_k \in P$, $k\geq1$, denote by
  $S:=R_{r_{s}}^{D_{s}}(p_1,\dots,p_k)$ and define a mapping
  $f: S\cup P_{s}(p) \to S\cup P_{s}(p')$ such that
  $f(e)=\iota(e)$ for $e\in P_{s}(p)$ and $f(e)=e$ otherwise.
  If $p\not\in R_{r_{s+1}-r_{s}}^{D_{s+1}}(p_1,\dots,p_k)$, then
\smallskip
  $f$ is a color-preserving isomorphism between the induced subposets
  $\PPP[S\cup P_{s}(p)]$ and $\PPP[S\cup P_{s}(p')]$.
\end{corollary}
\begin{figure}[tb]
$$
\begin{tikzpicture}[scale=0.5]
\draw[thick] (2.1,1) ellipse (20mm and 12mm);
\draw[thick] (2.1,6) ellipse (20mm and 12mm);
\draw[thick] (11,3.5) ellipse (27mm and 12mm);
\draw[thick,color=gray,rotate=-13] (5.55,6.2) ellipse (71mm and 21mm);
\draw[thick,color=gray,rotate=13] (7.15,0.6) ellipse (71mm and 21mm);
\draw (0.1,1) node[label=left:$P_{s}(p)$] {};
\draw (0.1,6) node[label=left:$P_{s}(p')$] {};
\draw (6.2,3.2) node[label=right:$f$] (f) {};
\draw (-0.5,3.2) node[label=left:$\iota\!\!$] (iota) {};
\draw[color=gray,line width=1.5mm,->] (0.5,2.1) .. controls (iota) .. (0.5,5);
\draw[color=gray,line width=3mm,->] (6,1.5) -- (6,5.5);
\tikzstyle{every node}=[draw, shape=circle, minimum size=2pt,inner sep=1.5pt,fill=yellow]
\draw (2,1) node[fill=blue, label=left:$p$~~~] (p) {}; 
\draw (2,6) node[fill=blue, label=left:{$p'$~~~}] (pp) {};
\draw (10,3.5) node[label=right:$\cdots\cdots$, label=left:$p_1$] (x) {};
\draw (12.2,3.5) node[label=right:$p_k$~~~~~~$S$] (y) {};
\tikzstyle{every path}=[->]
\draw (pp) -- (p);
\draw[color=gray,dashed] (x) -- (p); \draw[color=gray,dashed] (p) -- (y);
\draw[color=gray,dashed] (x) -- (pp); \draw[color=gray,dashed] (pp) -- (y);
\end{tikzpicture}
$$
\caption{An illustration of the statement of
Corollary~\ref{cor:isomorphic-neighborhoods}:
$f$ is a color-preserving isomorphism between the induced subposets
  $\PPP[S\cup P_{s}(p)]$ and $\PPP[S\cup P_{s}(p')]$.}
\label{fig:isomorphic-neighborhoods}
\end{figure}
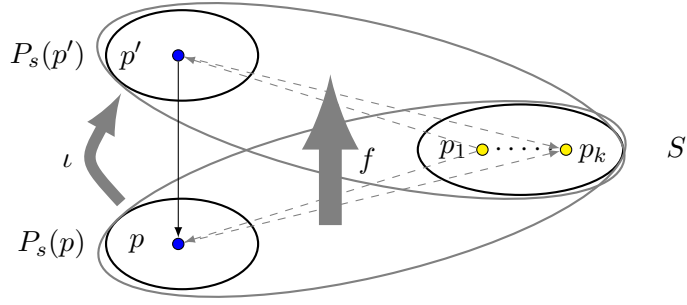

The statement of Corollary~\ref{cor:isomorphic-neighborhoods} is illustrated
in Figure~\ref{fig:isomorphic-neighborhoods}.

\begin{proof}
  For the sake of contradiction assume that $p\not\in
   R_{r_{s+1}-r_{s}}^{D_{s+1}}(p_1,\dots,p_k)$ and $f$ is not a poset
  isomorphism. Then either $f$ is not a bijection or there are two
  elements $a,b \in S\cup P_{s}(p)$ such that $a$ and $b$ are in a different
  relation with respect to $\leq^\PPP$ than $f(a)$ and $f(b)$.

  Observe that in the latter case, since  $a$ and $b$ are in a different relation with respect to
  $\leq^\PPP$ than $f(a)$ and $f(b)$, and by the definition of~$f$
  we clearly see that, up to symmetry,
  $a \in S \setminus P_{s}(p)$ and $b \in P_{s}(p)$.
  Hence, the latter case means that $a=f(a)$ discerns $b$ from $f(b)=\iota(b)$.

  In the former case consider the function $f'$
  defined analogously to the function $f$ but from the ``perspective of $p'$'',
  i.e., $f' : S \cup P_s(p') \rightarrow S \cup P_s(p)$ is defined by setting
  $f'(e)=\iota^{-1}(e)$ for every $e \in P_s(p')$ and $f'(e)=e$
  otherwise. Then $f$ is a bijection if and only if $f'$ is.
  Moreover, $f$ is a bijection if and only if both $f$ and $f'$ are injective. 
  If $f$ is not injective, then there exists a 
  $b \in P_{s}(p)$ such that $a=f(b)=\iota(b) \in S \setminus P_{s}(p)$.
  Since $a\not=b$, the element $a$ discerns $b$ from $\iota(b)=a$.
  On the other hand, if $f'$ is not injective, then there exists a
  $b' \in P_{s}(p')$ such that $a'=f'(b')=\iota^{-1}(b') \in S \setminus
  P_{s}(p')$.
  Since $a'\not=b'$, the element $a'$ discerns $a'$ from $b'$.
  Setting $a:=a'$ and $b:=a'$,
  we again see that $a$ discerns $b=a'$ from $\iota(b)=b'$.

  In either of the three subcases above, the elements $a$ and $b$ satisfy the conditions of Lemma~\ref{lem:in-neighborhood}.
  Hence there exists
  a directed path from $a$ to an element $a' \in C(p)$
  of type $\tau_{s+1}(p)$ with $p \leq^\PPP\! a' \leq^\PPP\! p'$
  in $D_{s}$ of length at most $2r_{s}+1$.
  Because of Lemma~\ref{lem:dgraph-edge-preservation}
  this path exists also in $D_{s+1}$. Furthermore, since $(p',p)$ is an arc
  with label $\eldown$ in $D_{s+1}$, there are no
  elements of type $\tau_{s+1}(p)$ on $C(p)$ ``between'' $p$ and $p'$,
  and so either $a'=p$ or $a'=p'$.
  Hence, there is a directed path from $a$ to
  $p$ in $D_{s+1}$ of length at most $2r_{s}+1+1$.

  Finally, since $a \in S$, there is a directed path from some
  of the elements $p_1, \dotsc, p_{k}$ to $a$ of length at most $r_{s}$.
  In a summary, there exists a directed path in $D_{s+1}$ from one of
  $p_1, \dotsc, p_{k}$ to $p$ of length at most (recall $r_s=3\cdot 4^s-1$)
$$
    2r_{s}+2+r_{s}=(4r_{s}+3)-r_{s}-1=r_{s+1}-r_{s}-1
$$
which contradicts the assumption $p\not\in R_{r_{s+1}-r_{s}}^{D_{s+1}}(p_1,\dots,p_k)$.

\smallskip
Color-preservation by $f$ immediately follows from the same property
of~$\iota$.
\end{proof}

\section{The Model Checking Algorithm}
\label{sec:FOmodelcheck}

We would like to use the Hintikka game to solve the
poset FO model checking problem $\PPP\models\phi$, by
Proposition~\ref{prop:Hintikka}.
Though, the number of possible distinct plays in the game $\cG(\PPP,\phi)$
grows roughly as $\bigO(\Vert\PPP\Vert^{|\phi|})$ which is not FPT.
To resolve this problem, we are going to show that in fact only a small
subset of all plays of the game $\cG(\PPP,\phi)$ is necessary to determine
the outcome---only those plays which are suitably locally constrained with the use of
Definition~\ref{def:graphDs}.

Let $\PPP=(P,\leq^\PPP)$ be a poset, with an implicitly associated 
coloring $\lambda:P\to\Lambda$ as in Section~\ref{sec:poset-types},
and $\phi$ an FO formula.
The {\em$r$-local Hintikka game} $\cG_r(\PPP,\phi)$
(where ``$r$'' refers to the sequence $r_s=3\cdot 4^s-1$,
cf.~Definition~\ref{def:graphDs})
is played on the same set of positions by the same rules as the ordinary
Hintikka game $\cG(\PPP,\phi)$, with the following additional restriction:
for each $Q \in \{\exists, \forall\}$, and for any position of the form
$\psi(p_1,\dotsc,p_{i}) \equiv Q y.\, \psi'(p_1,\dotsc,p_{i},y)$ where $i\geq1$,
Player $Q$ has to move to a position $\psi'(p_1,\dotsc,p_{i},p)$
such that $p\in R_{r_{q}-r_{q-1}}^{D_{q}}(p_1, \dotsc, p_{i})$,
where $q\geq1$ is the quantifier rank of $\psi'$.
If $q=0$, i.e., for quantifier-free $\psi'$, the restriction is
$p\in R_{r_0}^{D_0}(p_1, \dotsc, p_{i})$.

\begin{LEM}\label{lem:local-is-local}
Let $\PPP$ be a poset, $p_1,\dotsc,p_{j}\in P$, and $\phi$ be an FO formula
of quantifier rank~$q$.
In the $r$-local Hintikka game $\cG_r(\PPP,\phi)$ with an initial position
$\phi(p_1,\dotsc,p_{j})$, $j\geq1$, every reachable game position
$\psi(p_1,\dotsc,p_{k})$, $k>j$, is such that
$$p_{j+1},\dots,p_k\in R_{r_{q-1}}^{D_{q-1}}(p_1, \dotsc, p_{j}) .$$
\end{LEM}
\begin{proof}
We proceed by induction on $(k-j)$, starting with the trivial degenerate
case~$k=j$ and proving a stronger statement
$$p_{j+1},\dots,p_k\in R_{r_{q-1}-r_{q+j-k-1}}^{D_{q-1}}(p_1, \dotsc, p_{j}) .$$
Let the claim hold for a position $\psi(p_1,\dotsc,p_{k-1})$ and consider
a next position $\psi'(p_1,\dotsc,p_{k-1},p_{k})$, where the quantifier rank of
$\psi'$ is~$q'=q-(k-j)$.
Then, by the definition, $p_k$ is at distance at most $r_{q'}-r_{q'-1}$
in $D_{q'}$ from one of the elements $p_1,\dotsc,p_{k-1}$, and the same
holds also in $D_{q-1}$ by Lemma~\ref{lem:dgraph-edge-preservation} as
$q'\leq q-1$.
Since each of $p_1,\dotsc,p_{k-1}$ is at distance at most 
$r_{q-1}-r_{q+j-(k-1)-1}=r_{q-1}-r_{q'}$ from one of $p_1,\dotsc,p_{j}$
(and this holds also for $k-1=j$),
we get that $p_k$ is at distance at most 
$r_{q-1}-r_{q'}+r_{q'}-r_{q'-1}=r_{q-1}-r_{q'-1}=r_{q-1}-r_{q+j-k-1}$
from one of $p_1,\dotsc,p_{j}$ in $D_{q-1}$.
\end{proof}

Now we get to the crucial technical claim of this paper.
For $Q \in \{\exists, \forall\}$, and
with a neglectable abuse of terminology, we say that Player $Q$
{\em wins} the (ordinary or $r$-local) Hintikka game {\em from a position}
$\psi(p_1,\dotsc,p_{i})$ if Player $Q$ has a winning strategy
in the game $\cG(\PPP,\psi)$ (or in $\cG_r(\PPP,\psi)$, respectively)
with the initial position $\psi(p_1,\dotsc,p_{i})$.
Otherwise, Player $Q$ {\em loses} the game.

\begin{LEM}\label{lem:game-to-local-game}
  Let $\PPP=(P,\leq^\PPP)$ be a colored poset and $\phi$ be an FO formula in
  negation normal form.
  For $Q\in \{\exists, \forall\}$ and $i\geq0$, consider a position
  $\psi(p_1,\dotsc,p_{i})$
  in the Hintikka game $\cG(\PPP,\phi)$. 
  If Player $Q$ wins the game $\cG(\PPP,\phi)$
  from the position $\psi(p_1,\dotsc,p_{i})$,
  then Player $Q$ wins also the $r$-local Hintikka game $\cG_r(\PPP,\psi)$
  from an initial position $\psi(p_1,\dotsc,p_{i})$.
\end{LEM}

\begin{proof}
Let $\bar Q$ denote the other player, that is, $\{Q,\bar Q\}=\{\exists,\forall\}$.
For the sake of contradiction, assume that we have got a counterexample
with $\PPP$, $\phi$, and $\psi(p_1,\dotsc,p_{i})$ of quantifier rank~$q+1$;
meaning that Player $Q$ wins $\cG(\PPP,\phi)$ from $\psi(p_1,\dotsc,p_{i})$ but $Q$
loses $\cG_r(\PPP,\psi)$ from initial $\psi(p_1,\dotsc,p_{i})$.
Assume, moreover, that $Q\in \{\exists, \forall\}$ and the counterexample
are chosen such that the pair $\langle q,|\psi|\rangle$ is lexicographically minimal.

Since a game move associated with a conjunction or disjunction of
formulas, and a (possible) initial move for $i=0$, are not in any way restricted in
the $r$-local Hintikka game, our minimality setup guarantees that
$\psi$ starts with a quantifier, and so $q+1\geq1$ and $i\geq1$.
If this leading quantifier of $\psi$ was $\bar Q$, then we would again get
a contradiction to the minimality of~$q$.
Therefore,
\begin{equation}\label{eq:psi-to-psi'}
\psi(p_1,\dotsc,p_{i}) \equiv Q x_{i+1}.\,\psi'(p_1,\dotsc,p_{i},x_{i+1})
.\end{equation}

Note that $q$ is the quantifier rank of $\psi'$.
If $q=0$, then $\psi'$ is actually quantifier-free and Player $Q$ can make his
move $x_{i+1}$ with the element $p'\in R_{r_0}^{D_0}(p_1,\dotsc,p_{i})$,
$r_0=2$, as in Lemma~\ref{lem:equivalent-element-in-neighborhood}.
Since this is a contradiction to Player $Q$ losing $\cG_r(\PPP,\psi)$ from
$\psi(p_1,\dotsc,p_{i})$, we may further assume that $q\geq1$.

Let $p\in P$ be such that Player $Q$ wins the game $\cG(\PPP,\phi)$
from \eqref{eq:psi-to-psi'} $\psi(p_1,\dotsc,p_{i})$ by moving to the position
$\psi'(p_1,\dotsc,p_{i},p)$, and assume that $p$ is chosen maximal with
respect to $\leq^\PPP$ with this property.
By our minimal choice of $q$, the statement of
Lemma~\ref{lem:game-to-local-game} holds for the game $\cG(\PPP,\phi)$
from the position $\psi'(p_1,\dotsc,p_{i},p)$,
and so 
\begin{equation}\label{eq:local-win-Q}
\parbox{.6\hsize}{Player $Q$ wins the $r$-local game $\cG_r(\PPP,\psi')$ from
the initial position $\psi'(p_1,\dotsc,p_{i},p)$.}
\end{equation}
Consequently, 
\begin{equation}\label{eq:p-away}
p\not\in R_{r_{q}-r_{q-1}}^{D_{q}}(p_1, \dotsc, p_{i})
\end{equation}
since, otherwise, Player~$Q$ would win also the $r$-local game $\cG_r(\PPP,\psi)$ 
from $\psi(p_1,\dotsc,p_{i})$ which is not the case by our assumption.

Let $p' \in C(p)$ be the bottommost element such that $p\lneq^P\! p'$
and $\tau_{q}(p')=\tau_{q}(p)$.
Observe that such $p'$ does exist since, otherwise, $D_{q}$ would contain an arc
from the topmost element of $C(p)$ to $p$ with label \eldown, and hence
$p \in R_2^{D_{q}}(p_1)$ contradicting \eqref{eq:p-away}.
By our maximal choice of $p$ with respect to~$\leq^\PPP$,
Player $Q$ loses the game $\cG(\PPP,\phi)$
from $\psi(p_1,\dotsc,p_{i})$ after moving to the
position $\psi'(p_1,\dotsc,p_{i},p')$.
In other words, Player $\bar Q$ wins the game $\cG(\PPP,\phi)$
from the position $\psi'(p_1,\dotsc,p_{i},p')$ and, by our choice
of a counterexample with minimum~$q$;
\begin{equation}\label{eq:local-win-barQ}
\parbox{.6\hsize}{Player $\bar Q$ wins the $r$-local game $\cG_r(\PPP,\psi')$ from
the initial position $\psi'(p_1,\dotsc,p_{i},p')$.}
\end{equation}

Now, we employ Corollary~\ref{cor:isomorphic-neighborhoods}
for $s=q-1$, $k=i$ and $S=R_{r_{q-1}}^{D_{q-1}}(p_1,\dots,p_i)$.
By \eqref{eq:p-away}, we hence get that there exists a color-preserving isomorphism map
$f$ between the induced colored subposets $\PPP[S\cup P_{q-1}(p)]$ and 
$\PPP[S\cup P_{q-1}(p')]$.
By Lemma~\ref{lem:local-is-local},
all the elements $e$ played in any play of an $r$-local Hintikka game
$\cG_r(\PPP,\psi')$ from an initial position $\psi'(p_1,\dotsc,p_{i},p')$
belong to the set $R_{r_{q-1}}^{D_{q-1}}(p_1,\dots,p_i,p')=S\cup P_{q-1}(p')$.

The latter finding implies that every play of the $r$-local Hintikka game
$\cG_r(\PPP,\psi')$ from $\psi'(p_1,\dotsc,p_{i},p')$
can be duplicated, with the same outcome,
in the same game from $\psi'(p_1,\dotsc,p_{i},p)$ via the isomorphism
map~$f^{-1}$.
Therefore, by \eqref{eq:local-win-barQ}, Player $\bar Q$ wins the $r$-local game 
$\cG_r(\PPP,\psi')$ from the initial position $\psi'(p_1,\dotsc,p_{i},p)$, too.
However, this contradicts \eqref{eq:local-win-Q},
and so there cannot be a counterexample to the statement of the lemma.
\end{proof}

\begin{REM}
Notice that it is actually not necessary to explicitly
use Lemma~\ref{lem:equivalent-element-in-neighborhood} in the proof of
Lemma~\ref{lem:game-to-local-game}\,---with a slightly modified setting this
base case comes out ``for free''.
Though, we think that the current proof with
Lemma~\ref{lem:equivalent-element-in-neighborhood} is easier to read and
to understand.
\end{REM}

We can now easily formulate and prove the main result:

\begin{THM}\label{thm:main}
  Let $\PPP=(P,\leq^\PPP)$ be a poset, associated with
  $\lambda:P\to\Lambda$ where $\Lambda$ is a finite set of colors,
  and let $\phi$ be an FO sentence in negation normal form.
  The existential player has a winning strategy in the $r$-local
  Hintikka game $\cG_r(\PPP, \phi)$ if, and only if, $\PPP \models \phi$.
\end{THM}
\begin{proof}
By Proposition~\ref{prop:Hintikka}, it is enough to prove the following:
\begin{itemize}\parskip-1pt\item
  Player $Q$, where $Q\in \{\exists, \forall\}$, has a winning strategy in
  the Hintikka game $\cG(\PPP, \phi)$ if, and only if,
  Player $Q$ has a winning strategy in the $r$-local
  Hintikka game $\cG_r(\PPP, \phi)$.
\end{itemize}
The ``only if'' direction follows from Lemma~\ref{lem:game-to-local-game}
for~$i=0$, and so we deal with the ``if'' direction.
Assume that we have a counterexample minimizing the
quantifier rank of~$\phi$, i.e., that Player $Q$ wins $\cG_r(\PPP, \phi)$
but does not win $\cG(\PPP, \phi)$.

If the only quantifiers in $\phi$ are $Q$, then the $r$-local winning
strategy of Player $Q$ can be used ``as is'' in the game $\cG(\PPP, \phi)$.
Hence, denoting by $\bar Q$ the opposite quantifier to $Q$,
the sentence $\phi$ contains a subformula $\psi\equiv \bar Q y.\,\psi'(y)$,
such that all occurrences of the quantifier $\bar Q$ in $\phi$ are contained
within~$\psi$.
Let $\psi(p_1,\dots,p_i)$ be a position reached by any winning strategy
of Player $Q$ in the $r$-local game $\cG_r(\PPP, \phi)$.
Since Player $Q$ loses $\cG(\PPP, \phi)$ in general, Player $\bar Q$ wins
the ordinary game $\cG(\PPP, \phi)$ from the position $\psi(p_1,\dots,p_i)$.

By Lemma~\ref{lem:game-to-local-game} again, Player $\bar Q$ hence wins
the $r$-local game $\cG_r(\PPP, \psi)$ from the initial position $\psi(p_1,\dots,p_i)$.
However, this contradicts the assumption that the strategy of Player $Q$
in $\cG_r(\PPP, \phi)$
containing the position $\psi(p_1,\dots,p_i)$ is winning for Player~$Q$.
\end{proof}

The following is then a straightforward observation.

\begin{COR}\label{cor:main}
  For a poset $\PPP$ let $\mathcal{T}_i(\PPP)$ denote the set of 
  types $\tau_i$ of rank $i$ occurring in $\PPP$, i.e. 
  $\mathcal{T}_i(\PPP) := \{\tau_i(p): p \in P\}$,
  where $\tau_i$ is as in Definition~\ref{def:graphDs}.
  Assume $\PPP^1,\PPP^2$ are colored posets and $\phi$ an FO sentence of quantifier
  rank $q$ in negation normal form.
  If $\mathcal{T}_{q-1}(\PPP^1)=\mathcal{T}_{q-1}(\PPP^2)$, then it holds;
  $\PPP^1\models\phi$ if and only if $\PPP^2\models\phi$.
\end{COR}
\begin{proof}
Let the types of elements of $\PPP^1$ and $\PPP^2$ be denoted by
$\tau_i^1$ and $\tau_i^2$, respectively, and the associated digraphs
by $D_i^1$ and $D_i^2$.

From Theorem~\ref{thm:main}, it is enough to prove that the existential
player wins the $r$-local game $\cG_r(\PPP^1, \phi)$ if, and only if,
he wins $\cG_r(\PPP^2, \phi)$.
For that we have to analyze the game up to the first quantifier move,
and since the starting conjunction or disjunction moves are irrelevant for
this analysis, we may without loss of generality assume that
$\phi\equiv Q y.\, \psi(y)$ where $Q\in \{\exists, \forall\}$.

If the claim (that the existential player wins $\cG_r(\PPP^1, \phi)$ iff
he wins $\cG_r(\PPP^2, \phi)$\,) is not true,
then Player $Q$ wins precisely one of the two considered games, say,
the $r$-local game $\cG_r(\PPP^1, \phi)$.
Let a winning move of Player $Q$ in $\cG_r(\PPP^1, \phi)$ be the 
position~$\psi(p_1)$ where $p_1\in P^1$.
By the assumption $\mathcal{T}_{q-1}(\PPP^1)=\mathcal{T}_{q-1}(\PPP^2)$,
there hence exists $p_2\in P^2$ such that $\tau_{q-1}^1(p_1)=\tau_{q-1}^2(p_2)$.
Let the latter be witnessed by an isomorphism
$\iota: P^1_{q-2}(p_1) \to P^2_{q-2}(p_2)$ between the structures
$\AAA^{\PPP^1}_{q-2}(p_1)$ and $\AAA^{\PPP^2}_{q-2}(p_2)$
which is, in particular,
a poset isomorphism between the
induced colored subposets $\PPP^1[P^1_{q-2}(p_1)]$ and $\PPP^2[P^2_{q-2}(p_2)]$.

By Lemma~\ref{lem:local-is-local}, for the formula $\psi$ of quantifier rank
$q-1$ and for $j=1$, every reachable position in the $r$-local game 
$\cG_r(\PPP^1,\phi)$ belongs to $P^1_{q-2}(p_1)$.
Consequently, Player $Q$ could win also the $r$-local game
$\cG_r(\PPP^2,\phi)$, by copying his winning strategy under the isomorphism
map~$\iota$.
This contradiction finishes the proof.
\end{proof}

\begin{THE}\label{thm:algorithm}
  Let $\PPP=(P,\leq^\PPP)$ be a poset of width~$w$,
  with elements colored by $\lambda:P\to\Lambda$ where $\Lambda$ is a finite set,
  and let $\phi$ be an FO sentence in negation normal form.
  There is an algorithm which decides whether $\PPP \models \phi$ in FPT
  time $f(w,\phi)\cdot\Vert\PPP\Vert^2$.
\end{THE}
\begin{proof}
According to Corollary~\ref{cor:main}, it is enough to know the set
$\mathcal{T}_{q-1}(\PPP)$ in order to decide whether $\PPP \models \phi$.
We thus proceed the algorithm in two steps:
\begin{enumerate}\parskip-1pt
\item\label{it:setT}
  We compute the set $\mathcal{T}_{q-1}(\PPP)$ (of rank-$(q-1)$ types).
\item\label{it:models}
  We decide whether $\PPP \models \phi$ using the set from Step~\ref{it:setT}.
\end{enumerate}

First of all, we show that the set of all possible types of a given rank is
finite if the poset width $w$ is bounded.
Since each type is a (sub)digraph of bounded radius, it is enough to argue that
the out-degrees in $D_i$ are bounded.
Indeed, by induction, the outdegree in $D_0$ is $4w|\Lambda|$.
In $D_{i+1}$, the outdegree is bounded from above by a function of $w$ and the
number of all possible types of rank $i+1$, which is finite by the inductive
assumption for $D_i$.

Therefore, Step~\ref{it:models} is a finite problem and we may decide
whether $\PPP \models \phi$ by a brute-force evaluation of $\phi$ on each
member of $\mathcal{T}_{q-1}(\PPP)$.
This takes time $f'(w,\phi)$.
As for Step~\ref{it:setT}, we start with computing a chain partition of
width~$w$, in time $g(w) \cdot \Vert\PPP\Vert^2$ by Proposition~\ref{pro:comp-chain-part}.
We then proceed exactly along the iterations of constructive
Definition~\ref{def:graphDs}.
Since the number of possible types is finite, this computation takes time
at most $f''(w,\phi)\cdot\Vert\PPP\Vert^2$;
by traversing, in every iteration $D_i$ for $i=0,1,\dots,q-2$, 
for each $p\in P$ every chain of $\PPP$ and finding the appropriate out-neighbors.
\end{proof}

\section{Application to Interval Graphs}\label{sec:interval}

Besides the very successful story of FPT~ FO model checking on sparse graph
classes, culminating with the ultimate and outstanding result of Grohe,
Kreutzer, and Siebertz~\cite{gks14},
only a few such results have been published for dense graph classes
(especially, for graph classes which cannot be easily interpreted
in nowhere dense classes).
One of such notable papers is \cite{GHKOST13}, dealing with
FO model checking on interval graphs.

It has been shown \cite{GHKOST13} that FO model checking of interval graphs
is FPT when the intervals are restricted to have lengths from a fixed 
finite set of reals (Corollary~\ref{cor:INT-L}),
while the problem is W-hard whenever the intervals are allowed to have
lengths from any dense subset of a positive-length interval of reals.
We will demonstrate the strength and usefulness of our main result by giving a
rather short and straightforward derivation of the FPT result of
\cite{GHKOST13} from our Theorem~\ref{thm:algorithm}.

A graph $G$ is an {\em interval graph} if there exists a set $\cal I$ of
intervals on the real line such that $V(G)=\cal I$ and $E(G)$ is formed by
the intersecting pairs of intervals.
For a set $L$ of reals, a set $\cal I$ of intervals is called an
{\em$L$-interval representation} if every interval from $\cal I$ has
its length in~$L$.
This notion generalizes well-studied {\em unit interval graphs}
(where all interval lengths are~$1$), which are also known under the name
of proper interval graphs.
A set $\cal I$ of intervals is called a {\em proper interval representation}
if there is no pair of intervals $J_1,J_2\in\cal I$ such that 
$J_1$ is strictly contained in $J_2$ ($J_1\subsetneq J_2$).
We call $\cal I$ a {\em$k$-fold proper interval representation}
if there exists a partition ${\cal I}={\cal I}_1\cup\dots\cup {\cal I}_k$
such that each ${\cal I}_j$ is a proper interval representation for
$j=1,\dots,k$.

\begin{THE}\label{thm:INT-k-proper}
Let $\phi$ be a graph FO sentence.
Assume $G$ is an interval graph given along with its 
$k$-fold proper interval representation $\cal I$.
Then the FO model checking problem $G\models\phi$, parameterized by $k$ and
$\phi$, is FPT.
\end{THE}

\begin{proof}
First of all, we prove that we can, without loss of generality,
assume that no two ends of intervals from $\cal I$ coincide.
This can be achieved by a tiny perturbation of the intervals, as follows.
Let the given interval representation of $G$ be 
${\cal I}=\{[a_i,b_i]: i=1,\dots,n\}$, where $n=|V(G)|$.
Let $D=\bigcup_{i=1}^n\{a_i,b_i\}$ be the set of all interval ends.
Choose $\varepsilon>0$ such that the least positive difference in the
set $D$ is greater than $2\varepsilon$, and define
$${\cal I}' := \{[a_i+\varepsilon\frac in, \>
			b_i+\varepsilon(1+\frac in)]: i=1,\dots,n\} .$$
Then, clearly, the intersection graph of $\cal I'$ is isomorphic to $G$,
no two interval ends in $\cal I'$ coincide, and two intervals of $\cal I'$
are in a proper inclusion only if the same is true already in~$\cal I$.

Let ${\cal I}={\cal I}_1\cup\dots\cup {\cal I}_k$ be such that each 
${\cal I}_j$ is a proper interval representation for $j=1,\dots,k$.
Let $P:=D\cup\cal I$.
We define a poset $\PPP=(P,\leq^\PPP)$ as follows:
\begin{itemize}
\item
for numbers $d_1,d_2\in D$ it is $d_1\leq^\PPP\!d_2$ iff $d_1\leq d_2$,
\item
for $j\in\{1,\dots,k\}$ and intervals $J_1,J_2\in{\cal I}_j$, it is
$J_1\leq^\PPP\!J_2$ iff $J_1$ is not to the right of $J_2$, and
\item
for every interval $J=[a,b]\in\cal I$ and every $d\in D$, it is
$J\leq^\PPP\!d$ iff $d\geq b$, and $d\leq^\PPP\!J$ iff $d\leq a$.
\end{itemize}
The set $P$ can be partitioned into $k+1$ chains;
$D$ and ${\cal I}_1,\dots,{\cal I}_k$.
Hence the width of $\PPP$ is at most $k+1$.

\smallskip
Finally, we show how the poset $\PPP$ can be used to decide whether
$G\models\phi$, applying a so called FO interpretation of $G$ in~$\PPP$.
Specifically, we construct a poset FO sentence $\psi$, depending only on $\phi$, 
such that $\PPP\models\psi$ if and only if $G\models\phi$.
The rest then follows from Theorem~\ref{thm:algorithm}.

Let $D(x)$ be a unary predicate (a color) on $\PPP$ identifying the elements
of $D\subseteq P$.
Then, the proposition $\neg D(x)$ simply identifies the domain 
(vertex set) of~$G$ within~$P$ in our interpretation.
The crucial part of our construction of $\psi$ is to express by an FO formula
$\beta(J_1,J_2)$ that arbitrary intervals $J_1,J_2\in{\cal I}$ 
intersect each other (i.e., form an edge of~$G$).
Note that if intervals $J_1,J_2$ do not intersect, 
then $J_2$ is to the right of the right end of $J_1$ or vice versa.
Then, by our definition of~$\PPP$,
$J_2$ is to the right of the right end of $J_1$ if, and only if,
there exists $d\in D$ such that $J_1\leq^\PPP\!d\leq^\PPP\!J_2$
($d$ lies ``between'' $J_1$ and $J_2$).
Hence we can express the edge relation of $G$ as
$$
\beta(x,y)\>\equiv\> \forall d \big[ D(d) \to \big(
		(\neg\, x\leq^\PPP\!d \vee \neg\, d\leq^\PPP\!y) \wedge
		(\neg\, y\leq^\PPP\!d \vee \neg\, d\leq^\PPP\!x)
	\big)\big]
,$$
meaning that there exists no $d\in D$ such that $d$ lies ``between'' the
intervals $x$ and~$y$ (or, $y$ and $x$) of~$\cal I$.

The construction of $\psi$ then proceeds by structural induction on the structure
of~$\phi$ as follows.
Each occurrence of atomic $edge(x,y)$ is replaced with $\beta(x,y)$.
Other atomic propositions and logic connectives are simply copied from
$\phi$ to $\psi$.
Each quantifier $\exists x\,\sigma(x)$ is replaced with
$\exists x(\neg D(x)\wedge\sigma(x))$, and
$\forall x\,\sigma(x)$ with $\forall x(\neg D(x)\to\sigma(x))$.
All this construction is achieved in polynomial time.
Based on previous arguments, it is routine to verify that 
$\PPP\models\psi$ $\iff$ $G\models\phi$.
\end{proof}

We can now match the main result of \cite{GHKOST13}, except the precise runtime:
\begin{COR}[Ganian et al.~\cite{GHKOST13}]\label{cor:INT-L}
For every finite set $L$ of reals,
the FO model checking problem of $L$-interval graphs
(given alongside with an $L$-interval representation)
is FPT when parameterized by the FO sentence $\phi$ and~$|L|$.
\end{COR}
\begin{proof}
Let $\cal I$ be a given $L$-interval representation, and set $k:=|L|$.
We partition ${\cal I}={\cal I}_1\cup\dots\cup {\cal I}_k$ such that
each ${\cal I}_i$ contains intervals of the same length (from~$L$),
and then apply Theorem~\ref{thm:INT-k-proper}.
\end{proof}

Although, the formulation of Theorem~\ref{thm:INT-k-proper} is more general
than~\cite{GHKOST13}.
We can, for instance, in the same way derive fixed-parameter tractability 
also for FO model checking of well-studied {\em$k$-proper interval graphs},
introduced in \cite{PT99} as those having an interval representation
such that no interval is properly contained in more than $k$ other intervals.

\section{Conclusions}\label{sec:concl}

Our result can be seen as an initial step towards an understanding of the
complexity of FO model checking on non-sparse classes of structures
and we hope that the techniques developed here will be
useful for future research in this direction, e.g.,
to investigate the complexity of FO model checking on other algebraic
structures such as finite groups and lattices as suggested
by Grohe~\cite{gks14}, as well as on other dense graph classes,
to which the established ``locality'' tools of finite model theory do not apply. 

The result 
may also be used directly towards establishing fixed-parameter tractability for FO model checking
of other graph classes. Given the ease with which it implies
the otherwise non-trivial result on interval graphs~\cite{GHKOST13}, it is a natural to ask which other (dense) graph classes
can be interpreted in posets of bounded width.

\begin{small}
\bibliographystyle{plain}
\bibliography{posets-fo}
\end{small}

\end{document}